\documentstyle[11pt]{article}

\begin{document}           
\title{
The Tolman-Bondi Model in the Ruban-Chernin Coordinates. 1.
Equations and Solutions.
\thanks{gr-qc/9612038}
}
\author{
Alexander Gromov \thanks{
St. Petersburg State Technical University,
Faculty of Technical Cybernetics, Dept. of Computer Science,
29, Polytechnicheskaya str. St.-Petersburg, 195251, Russia
and
Istituto per la Ricerca di Base,
Castello Principe Pignatelli del Comune di Monteroduni,
I-86075 Monteroduni(IS), Molise, Italia.
E-mail: gromov@natus.stud.pu.ru
}}
\maketitle
\begin{abstract}
The Tolman-Bondi (TB) model is defined up to some transformation of a
co-moving coordinate but the transformation is not fixed.
The use of an arbitrary co-moving system of coordinates leads to
the solution dependent on three functions $f, F, {\bf F}$ which are chosen
independently in applications.

The article studies the transformation rule which is given by the definition
of an invariant mass.
It is shown that the addition of the TB model
by the definition of the transformation rule leads to the separation of
the couples of functions ($f$, $F$) into nonintersecting
classes.
It is shown that every class is characterized only by the
dependence of $F$ on $f$ and connected with unique system of co-moving
coordinates. It is shown that
the Ruban-Chernin system of coordinates corresponds to identical
transformation.
The dependence of Bonnor's solution on the Ruban-Chernin coordinate $M$
by means of initial density and energy distribution is studied. It is shown
that the simplest flat solution is reduced to an explicit dependence on the
coordinate $M$.
Several examples of initial conditions and transformation rules are
studied.
\\
PACS number(s): 95.30.Sf, 98.65.-r, 98.80.-k
\\
key words: cosmology:theory --- gravitation --- relativity
\end{abstract}
%
\section{The Introduction} \label{Introduction} \label{s 1}

The observations show that in the
large scale the Universe is not homogeneous. At the same time it is also
supposed that the secreted centre is missing.
These two properties separately are presented in the
Friedmann-Robertson-Walker (FRW) and Tolman-Bondi (TB) models
(Tolman 1934; Bondi 1947):
the FRW
model is homogeneous and does not include the secreted centre; the TB is
nonhomogeneous and includes one.

As the simplest nonhomogeneous model the TB model is used for
interpretation of the observation data.
The TB model is used to calculate the
redshift (Bondi 1947; Moffat 1992; Ribeiro 1992a, 1993)
to describe
the local void (Moffat 1995; Bonnor \& Chamorro 1990; Chamorro 1991)
to
interpret the fractal structure of the matter distribution in the
Universe (Ribeiro 1992a; Ribeiro 1992b; Ribeiro 1993).
A place of the TB model among the models consistent with the modern
observation data is shown in (Baryshev et al. 1994).

The TB model (Bondi 1947; Tolman 1934)
describes the spherical symmetry dust motion
with zero pressure in a co-moving system of coordinates.
The solution of the Tolman's equations
obtained by Bonnor (1972; 1974)
contains two undetermined functions $f(r)$ and $F(r)$
of the co-moving coordinate $r$ which
are defined by an initial conditions of the model.
But the co-moving coordinate
is defined up to a continuous
transformation $r^{\prime} = \Phi (r)$, so the
functions $f(r)$ and $F(r)$ are defined nonuniquely. This makes difficulty
in an interpretation of the observation data.

The article is based on the Bonnor's solution and on the univalent
definition of the co-moving coordinate given by Ruban
and Chernin (1969).

The article is dedicated to classification of the transformation $\Phi$,
to formulation an initial conditions through density and energy
distribution, to representation of the TB model in the Ruban-Chernin
system of coordinates and to the study of some examples of function $f$ and
$F$ and initial conditions.

It is shown that the Bonnor's flat solution of the TB model
is reduced to an explicit dependence on the co-moving coordinate $M$
in the Ruban-Chernin system of coordinates.

The Newtonian analog of the TB model has been studied by
Ruban and Chernin (1969).

\section{The TB Model} \label{TB model}

The TB model
is represented by an interval, an equation of motion for the
metrical function $\omega(r,t)$ with initial conditions for it
and an equation for density (Tolman 1934).

In the spherical symmetry and co-moving system of coordinates
the interval of the TB model has the form
\begin{equation}
{\rm d}s^2(r,t) =
{\rm e}^{\omega(r,t)}\,\frac{\omega^{\prime 2}(r,t)}{4\,f^2(r)}
{\rm d}r^2 + {\rm e}^{\omega(r,t)}
\left(
{\rm d}\theta^2 + \sin ^2\theta {\rm d}\phi
\right) - c^2 {\rm d}t^2,
\label{ds2}
\end{equation}
where $c$ is the velocity fo light, $t$ is time,
$\omega(r,t)$ is the metrical function,
$f(r)$ is the undetermined function, $\dot {} = \frac{\partial}{c \partial
t}$.
The metric (\ref{ds2}) is defined up to a continuous
transformation $\Phi$ of the co-moving coordinate $r$ (Landau \& Lifshits
1973).

The system of Einstein's equation is reduced to the equation of motion
and the equation of density. The equation of motion is:
\begin{equation}
{\rm e}^{\omega(r,t)}
\left(
 \ddot \omega(r,t) +\frac{3}{4} \dot \omega^2(r,t) - \Lambda
\right)
+
\left[
1 - f^2(r)
\right] = 0
\label{T:10}
\end{equation}
with initial conditions
\begin{equation}
\left.\omega(r,t)\right|_{t = 0} = \omega_0(r), \quad
\left.\dot\omega(r,t)\right|_{t = 0} = \dot\omega_0(r),
\label{i-c}
\end{equation}
where $\Lambda$ is a cosmological constant
and $\omega_0(r)$ and $\dot\omega_0(r)$ are given functions.

Using the Bonnor (1972) notation
\begin{equation}
R(r,t) = {\rm e}^{\omega(r,t)/2},
\label{B}
\end{equation}
where $R(r,t)$ is an analog of the Euler coordinate of the particle with
co-moving coordinate $r$, we rewrite the initial conditions in the form
\begin{equation}
\left.R(r,t)\right|_{t = 0} = R_0(r) = {\rm e}^{\omega_0(r)/2},
\label{Bb-1}
\end{equation}
\begin{equation}
\left.\dot R(r,t)\right|_{t = 0} = \dot R_0(r) =
\frac{1}{2}\,\dot\omega_0(r)\,{\rm e}^{\omega_0(r)/2}. \nonumber
\label{Bb-2}
\end{equation}
First integral of the equation (\ref{T:10}) is:
\begin{equation}
\frac{1}{2}\left(\frac{\partial R(r,t)}{\partial t}\right)^2 =
c^2\,\frac{f^2(r) - 1}{2} +
c^2\,\frac{F(r)}{4\,R(r,t)} +
\frac{\Lambda}{6}\,c^2\,R^2(r,t),
\label{TB-klass}
\end{equation}
where $F(r)$ is the second undetermined function.
The function $F(r)$ is defined from (\ref{Bb-1}), (\ref{Bb-2}) and
(\ref{TB-klass}).
If $\Lambda = 0$ then the equation (\ref{TB-klass}) may be interpreted as
an analog of the energy conservation law (Bondi 1947).

The integral of the equation (\ref{TB-klass}) has the form
\begin{equation}
\pm t + {\bf F}(r) =
\int\limits_{R_0(M)}^{R(M,t)}
\frac{{\rm d} \tilde R}{
\sqrt{
f^2(r) - 1 + \frac{F(r)}{2\,\tilde R} + \frac{\Lambda}{3}\tilde R^2}},
\label{T:15:11}
\end{equation}
where ${\bf F}(r)$ is the third undetermined function, the sign $'+'$
corresponds to an expanding solution and the sign $'-'$ corresponds to the
falling one.

The definition of density is given in the TB model by the
formula
\begin{equation}
\frac{8 \pi G}{c^2} \rho(r,t) = \frac{{\rm d} F(r)}{{\rm d}
r}\,\frac{1}{2\,R^2(r,t)\,\displaystyle\frac{\partial R(r,t)}{\partial r}}.
\label{T:15:1}
\end{equation}

\section{The Transformation Rules of the Co-moving Coordinates
and the Ruban-Chernin System of Coordinates} \label {s 3}

In accordance with (\ref{T:15:1}) the invariant density has the form
\begin{equation}
\rho(r,t)\,\sqrt{-g(r,t)} =
\frac{c^2}{16\,\pi\,G} \frac{F^{\prime}(r)}{f(r)}
\label{i-d}
\end{equation}
and does not depend on time. The invariant mass is
$$M(r) = 4\,\pi\,\int\limits_{0}^{r}\rho(r,t)\,\sqrt{-g}\,{\rm d}r =$$
\begin{equation}
4\,\pi\,\int\limits_{0}^{r}\rho(r,t)
\frac{R^2(r,t)}{f(r)}\frac{\partial R(r,t)}{\partial r}
\,{\rm d}r =
\frac{c^2}{4\,G}
\int\limits_{0}^{r}
\frac{F^{\prime}(r)}{f(r)}
{\rm d} r.
\label{Bondi}
\end{equation}
(\ref{Bondi}) has the sense of transformation rule from $r$ to $M$
and is specified by the function
\begin{equation}
\Psi(r) = \frac{c^2}{4\,G}\,\frac{F^{\prime}(r)}{f(r)}.
\label{Psi}
\end{equation}

The transformation (\ref{Bondi}) is read now
\begin{equation}
M(r) = \int\limits_{0}^{r}\Psi(\tilde r)\,{\rm d}\tilde r.
\label{newM}
\end{equation}
We suppose also that $M^{-1}$ exists and is the only one.
The uniqueness is
connected to the condition of the particle layers nonintersecting.
From (\ref{newM}) it follows that the transformation rule from some
coordinate $r_1$ to $r_2$ one is as follows:
\begin{equation}
\int\limits_{0}^{r_1}\Psi_1(r)\,{\rm d} r =
\int\limits_{0}^{r_2}\Psi_2(r)\,{\rm d} r.
\label{tr}
\end{equation}
The transformation (\ref{newM}) introduces a classification of the co-moving
coordinates and the functions $f$ and $F$.
The function $\Psi$ defines: 1) the transformation $r \to M$ and
position of the system coordinates $\{r,t\}$ with respect to the system
coordinates $\{M,t\}$;
2)the classes of the couples of the functions
$f(r)$ and $F(r)$ which are indiscernibled form the point of view of
the transformation rule (\ref{Bondi}); 3) the dependence between the
functions $f(r)$ and $F(r)$ inside every class:
\begin{equation}
F(r) = \frac{4\,G}{c^2}\,\int\limits_{0}^{r}\Psi(\tilde r)
\,f(\tilde r)\,{\rm d} \tilde r.
\label{F(f)}
\end{equation}
From (\ref{newM}) it follows that the classes of couples of the functions
$f$ and $F$ with different functions $\Psi$ are nonintersecting.

The transformation
\begin{equation}
\Phi: r \to  \tilde r
\label{rr}
\end{equation}
between two arbitrary co-moving coordinates $r$ and $\tilde r$
has been introduced in the section \ref{Introduction}. We can define it now
as
\begin{equation}
\Phi = \psi \, \circ  \, \tilde \psi^{-1},
\label{rr1}
\end{equation}
where
\begin{equation}
\psi: r \to M,\qquad \tilde \psi^{-1}: M \to \tilde r.
\label{MMrr}
\end{equation}
(\ref{rr1}) demonstrates that the transformation $\Phi$ between arbitrary
$r$ and $\tilde r$ is represented as the superposition of two
transformations having the form (\ref{newM}). So, the study
of the transformation (\ref{rr}) is reduced to the studying the
transformation (\ref{newM}) or, which is the same, the function $\Psi$.

There are three functions $f(r)$, $F(r)$ and $\Psi(r)$, only two of which
are
independent. The function $f(r)$ defines the metric (\ref{ds2}) and must be
used as an independent function.
The function $\Psi(r)$ marks the class of the transformation rules.
So, we can use the set of functions $\Psi$, $f$, ${\bf F}$ instead of
the set $f$, $F$, ${\bf F}$.

The function $F(r)$ is used as initial condition in a number of articles.
So, the use $F(r)$ or $\Psi(r)$ as independent function follows from
the context of the problem.

We define the Ruban-Chernin system of coordinates as the identical
transformation (\ref{newM}):
\begin{equation}
M = r.
\label{R-Ch-tr}
\end{equation}
The function $\Psi$ corresponding to this transformation is
\begin{equation}
\Psi(r) = 1
\label{R-Ch-tr-Psi}
\end{equation}
so, the dependence between $F(r)$ and $f(r)$ in case of identical
transformation is
\begin{equation}
F(M) = \frac{4\,G}{c^2}\,\int\limits_{0}^{M}f(\tilde M)\,{\rm d}\tilde M.
\label{F def}
\end{equation}

\section{The Definition of the Functions $f$, $F$ and ${\bf F}$
in the Ruban-Chernin Coordinates} \label{s 4}

First interpretation of the functions $f(r)$ and $F(r)$ has been presented
in (Bondi 1947). Following Bondi
let us compare the equation (\ref{TB-klass})
and the equation of the total energy in the Newtonian theory:
\begin{equation}
\frac{1}{2}\left(\frac{\partial R(r,t)}{\partial t}\right)^2 =
E(m) + \frac{G\,m}{R(m,t)},
\label{Newt}
\end{equation}
where $m$ is the mass of the sphere where the particle is located, $E(m)$
is the full specific energy.
In case of $\Lambda = 0$, the equations have the unique structure
concerning $R$, so
\begin{equation}
E_0(r) = c^2\,\frac{f^2(r) -1}{2}
\label{1-Newt}
\end{equation}
may be interpreted as an analog of a full specific energy and
\begin{equation}
\frac{c^2}{4 R(r,t)}\,F(r) = \frac{G\, m(r)}{R(r,t)}
\label{2-Newt}
\end{equation}
may be interpreted as
an analog of a specific potential energy, $m(r)$ is an effective mass
which will be defined.

Now let us study the case of two system of coordinates: a system of
an arbitrary coordinates
$\{r,t\}$ and the Ruban-Chernin system of coordinates $\{M,t\}$.
Suppose the functions $f(r)$ and $F(r)$ are specified.
The transformation rule from $\{r,t\}$ to $\{M,t\}$ is given by the formula
(\ref{Bondi}), where $M(r)$ is the continuous transformation up to
which the co-moving coordinate is defined.
So, we have three functions $f(r)$, $\Psi(r)$ and $M(r)$, any two of which
define the third function.

From (\ref{i-d}) and (\ref{F def}) it follows that the Ruban-Chernin
system of coordinates is singled out by the fact that the invariant
density is constant:
\begin{equation}
\rho(M,t)\,\sqrt{-g(M,t)} = \frac{1}{4\,\pi}.
\label{i-d-R-Ch}
\end{equation}
From (\ref{2-Newt}) it follows that the effective mass $m(M)$ is:
\begin{equation}
m(M) = \int\limits_{0}^{M}f(\tilde M)\,{\rm d} \tilde M.
\label{m(M) def}
\end{equation}
The initial condition $E_0(M)$ defines the function $f(M)$ by the
formula (\ref{1-Newt}), so the formula (\ref{F def}) becomes the definition
of the function $F(M)$.
The effective mass is equal to $M$ in case of $f = 1$.
We can now rewrite the equation (\ref{TB-klass}) in the form
\begin{eqnarray}
\frac{1}{2}\left(\frac{\partial R(M,t)}{\partial t}\right)^2 =
E_0(M) +
\frac{G}{R(M,t)}\,\int\limits_{0}^{M} \sqrt{1 + \frac{2}{c^2}\,E_0(M)}\,
{\rm d}M
+ \nonumber\\
\frac{\Lambda}{6}\,c^2\,R^2(M,t),
\label{TB-klass-n.u.}
\end{eqnarray}
where $$E_0(M) \geq -\frac{c^2}{2}. $$

Tolman (1934) also uses the third undetermined function
${\bf F}$ to solve the equation (\ref{TB-klass}).
The function ${\bf F}(M)$ in the Ruban-Chernin coordinates has the form
\begin{equation}
{\bf F}(M) =
\int\limits_{0}^{R_0(M)}
\frac{{\rm d} \tilde R}{
\sqrt{
2\,E_0(M) +2\,\frac{G\,m(M)}{\tilde R} + \frac{c^2\,\Lambda}{3}\tilde
R^2}}.
\label{T:12}
\end{equation}

\section{The Initial Conditions for the TB Model in the Ruban-Chernin
Coordinates} \label{s 5}

The equation of motion (\ref{T:10})
for metrical function $\omega(M,t)$ requires two initial conditions:
$\omega(M,0)$ and $\dot\omega(M,0)$. These functions are obtained in this
section.

Let us suppose that an initial profile of the density is given as the
function of the co-moving coordinate $M$. From (\ref{Bondi}) and
(\ref{R-Ch-tr}) it follows that
\begin{equation}
R(M,0) =
R_{0}(M) = \left[
\frac{3}{4\,\pi}\,\int\limits_{0}^{M}
\frac{f(\tilde M)}{\rho(\tilde M,0)}\,{\rm d} \tilde M
\right]^{1/3}.
\label{Mm}
\end{equation}
Substituting (\ref{B}) into (\ref{Mm}) we obtain the first initial
condition
\begin{equation}
\omega(M,0) = \frac{2}{3}\,\ln{\left[
\frac{3}{4\,\pi}\,
\int\limits_{0}^{M} \frac{f(\tilde M)}{\rho(\tilde M,0)}\,{\rm d} \tilde
M\right]}.
\label{1ic}
\end{equation}
The second initial condition is obtained by the substitution (\ref{B}),
(\ref{Bb-1}) and (\ref{Bb-2}) into (\ref{TB-klass-n.u.}):
\begin{eqnarray}
\left.
\left(
\frac{\partial \omega(M,t)}{\partial t}\right)^2\right|_{t=0} =
c^2\,\dot\omega^2(M,0) =
\nonumber\\
\\
 = 8\,\frac{E_0(M)}{R^2(M,0)} + 8\,\frac{G\,m(M)}{R^3(M,0)} +
\frac{4}{3}\,c^2\,\Lambda. \nonumber
\label{2ic}
\end{eqnarray}
For the initial profile of the velocity we obtain:
\begin{equation}
\frac{\partial R(M,0)}{\partial t} = \pm\sqrt{
2\,E_0(M) + 2\,\frac{G\,m(M)}{R(M,0)} + \frac{c^2\, \Lambda}{3}\,R^2(M,0)}.
\label{000}
\end{equation}
The equations (\ref{Mm}) - (\ref{000}) represent the initial conditions of
the TB model through the initial profiles of density and energy.
The last equation defines the initial profile of velocity.
The full specific energy $E_0(R)$ is limited by the meaning
$E_{min}(R)$ when $\frac{\partial R(M,0)}{\partial t} = 0$:
\begin{equation}
E_0(R) \ge E_{min}(R) = -\frac{G\,m(M)}{R(M,0)} -
\frac{c^2 \Lambda}{6}\,R^2(M,0).
\label{e_min}
\end{equation}

\section{The Initial Conditions for the FRW Model in the Ruban-Chernin
Coordi\-nates} \label{s 6}

The FRW model is the special case of the TB model which is specified
by the condition
\begin{equation}
\frac{\partial \rho(M,t)}{\partial M} = 0.
\label{FRW-1}
\end{equation}
So, only the first initial condition is changed and reads:
\begin{equation}
R_{FRW}(M,0) = \left[
\frac{3}{4\,\pi\,\rho_{FRW}(0)}
\int\limits_{0}^{M}
f(\tilde M)\,{\rm d} \tilde M
\right]^{1/3},
\label{FRW-2}
\end{equation}
\begin{equation}
\omega_{FRW}(M,0) = \frac{2}{3}\,\ln{\left[
\frac{3}{4\,\pi\,\rho_{FRW}(0)}\,
\int\limits_{0}^{M} f(\tilde M)\,{\rm d} \tilde M\right]}.
\label{FRW-3}
\end{equation}

\section{The Bonnor's Solution in the Ruban-Chernin Coordinates} \label{s 7}

The solution of the equation of the TB model with $\Lambda = 0$ has been
obtained by Bonnor (1972; 1974).
The solution of the Newtonian analog of the TB model has been obtained by
Ruban and Chernin and has the same form. We represent here the solution in
the Ruban-Chernin coordinates.

From the equation (\ref{TB-klass-n.u.}) it follows that
\begin{eqnarray}
R(M,t\,)\left(\frac{\partial R(M,t)}{\partial t}\right)^2 -
2\,E_0(M)\,R(M,t) = 2\,G\,m(M).
\label{TB-1u}
\end{eqnarray}
This equation is studied together with the initial conditions (\ref{Mm}).

The flat Bonnor's solution is obtained by the substitution of $E_0(M) = 0$
and brings to the equation
\begin{eqnarray}
R(M,t\,)\left(\frac{\partial R(M,t)}{\partial t}\right)^2 =
2\,G\,M
\label{TB-2u}
\end{eqnarray}
which has the solution
\begin{eqnarray}
R^{3/2}(M,t) = R_0^{3/2}(M) \pm 3\,\sqrt{\frac{G\,M}{2}}\,t,
\label{TB-3ugu}
\end{eqnarray}
satisfying the initial condition (\ref{Mm}).

In case of $E_0(M) \ne 0$ the solution has the following form depending on
the sign of the $E_0(M)$.
For $E_0(M) < 0$:
\begin{equation}
\left.
\begin{array}{c}
R(M,t) = R_m(M)\,\left\{1 - \cos [\eta(M,t)]\right\}, \\ \\
\pm\,t + {\bf F}(M) = \frac{R_m(M)}{\sqrt{2\,|E_0(M)|}}\,\left\{
\eta(M,t) - \sin [\eta(M,t)]
\right\} \\
\end{array}
\right\}
\quad E_0(M) < 0,
\label{E < 0}
\end{equation}
where
\begin{eqnarray}
R_m(M) &=& \frac{G\,m(M)}{|E_0(M)|}, \\
\eta(M,t) &=& 2\,\arcsin
\sqrt{\frac{R(M,t)}{R_m(M)}}.
\label{LTB-3uG}
\end{eqnarray}
In the Ruban-Chernin system of coordinates the function {\bf F}(M) has the
form:
\begin{eqnarray}
{\bf F}(M) = \frac{R_m(M)}{\sqrt{2\,|E_0(M)|}}
\left\{
\eta_0(M) - \sin [\eta_0(M)]
\right\},
\label{LTB-3u}
\end{eqnarray}
where
\begin{eqnarray}
\eta_0(M) = \eta(M,0)
\label{uuui}
\label{LTB-3ua}
\end{eqnarray}
In case of $E_0(M) > 0$:
\begin{equation}
\left.
\begin{array}{c}
R(M,t) = R_m(M)\,\left\{\cosh [\eta(M,t)] - 1\right\}, \\ \\
\pm\,t + {\bf F}(M) = \frac{R_m(M)}{\sqrt{2\,E_0(M)}}\,\left\{
\sinh [\eta(M,t)] - \eta(M,t)
\right\} \\
\end{array}
\right\}
\quad E_0(M) > 0,
\label{E > 0}
\end{equation}
where
\begin{eqnarray}
R_m(M) &=& \frac{G\,m(M)}{E_0(M)}, \\
\eta(M,t) &=& 2\,{\rm arcosh}
\sqrt{\frac{R(M,t)}{R_m(M)}}.
\label{LTB-3uF}
\end{eqnarray}
In the Ruban-Chernin system of coordinates the function {\bf F}(M) has the
form:
\begin{eqnarray}
{\bf F}(M) = \frac{R_m(M)}{\sqrt{2\,|E_0(M)|}}
\left\{
\sinh [\eta_0(M)] - \eta_0(M)
\right\},
\label{LTB-3uE}
\end{eqnarray}
where
\begin{eqnarray}
\eta_0(M) = \eta(M,0).
\label{uuui-1}
\end{eqnarray}

\section{The Examples of Functions $f(r)$, $F(r)$, $\Phi(r)$
and Initial Conditions} \label{s 8}

Up to this moment we have used one transformation $\Phi$, which has been
produced by the definition of the invariant mass. In Table 1 we
show several examples of co-moving coordinates given by Bonnor.

Table 2 shows the transformation rules from some system of
coordinates $\{r,t\}$ defined by two functions $f(r)$ and $F(r)$,
to the Ruban-Chernin system of coordinates $\{M,t\}$.
The transformation rule is produced by the given functions $f$ and $F$.

We will analyze now two examples of initial
conditions given by Bonnor (1974) and Ribeiro (1993).
They consider the flat Bonnor's solution in the form
\begin{eqnarray}
R(r,t) = \frac{1}{2}\,\left[9\,F(r)\right]^{1/3}\left[
t + \beta(r)
\right]^{2/3}.
\label{LTB-R15}
\end{eqnarray}
\begin{table}
\caption{
Examples of co-moving coordinates.
}
\begin{center}
\begin{tabular}{|c|c|}   \hline
author       & $\Phi: r \rightarrow \mbox{new coordinate}$ \\ \hline
Ruban & \\ and & \\ Chernin,1969 & $r \rightarrow M(r)$ \\
Bonnor, 1972 & $r \rightarrow \alpha \, R_0(r)$ \\
Bonnor, 1972 & $r \rightarrow \alpha \,\left(1 + k\,r^3\right)^{-1/6}$ \\
Bonnor, 1974 & $r \rightarrow \alpha\,F(r)$ \\
\hline
\end{tabular}
\end{center}
\end{table}
\begin{table}
\caption{
Examples of the functions $f, F, {\bf F}$ and transformation rules.
}
\begin{center}
\begin{tabular}{|c|c|c|c|c|}   \hline
author       & $f(r)$ & $F(r)$ & ${\bf F}$ & $M = \psi(r)$  \\ \hline
& & & & \\
Bonnor, 1972 & 1      & $\frac{\alpha\,R^{3}_0(r)}{\sqrt{1 +
k\,R^{3}_0(r)}}$ & $ \sim r^{3/2} $ & $4\,\pi\, \frac{\alpha\,R^{3}_0(r)}{\sqrt{1 +
k\,R^{3}_0(r)}}$ \\
Bonnor, 1974 & 1 & $\alpha\,r^3$ & $ {\bf F} = 0 $ &
$4\,\pi\,\alpha\,r^3$\\
Bonnor, 1974 & $\sqrt{1 + r^2}$ & $\alpha\,r^3$ & $r\,{\bf F^{\prime} \to 0}$ &
$6\,\pi\,\alpha\left(
r\,\sqrt{1+r^2} - {\rm arcsinh} (r)
\right)$ \\
Bonnor & & & & \\ and & & & & \\ Chamorro, 1991 &
$ \sin ^2(r)$ & $k\, \sin ^3(r)$  & $$ & $12\,\pi\,k\,{\rm sin}(r)$
\\
Ribeiro, 1993 & 1 & $\alpha\,r^p$  & $ \beta_0 + \eta_0\,r^q$ &
$4\,\pi\,\alpha\,r^p$ \\
Ribeiro, 1993 & $\cos (r)$ & $\alpha\,r^p$ & $ {\rm ln}({\rm e}^{\beta_0} +
\eta_1\,r)$ & $4\,\pi\,\alpha
\,p\,\int\limits_{0}^{r}\frac{\tilde r^{p-1}\,{\rm d}\tilde r}{\cos (\tilde
r)}$ \\
Ribeiro, 1993 & $\cosh (r)$ & $\alpha\,r^p$ & $ \beta_0 + \eta_0\,r^{q}
$ & $4\,\pi\,\alpha
\,p\,\int\limits_{0}^{r}\frac{\tilde r^{p-1}\,{\rm d}\tilde r}{\cosh (\tilde
r)}$ \\
\hline
\end{tabular}
\end{center}
\end{table}
Independently from the fact how the function $F$ depends on $r$, it
follows from (\ref{Bondi}) that in case of $f(r) = 1$
\begin{eqnarray}
F(M) = \frac{4\,G}{c^2}\,M.
\label{LTB-R2}
\end{eqnarray}
Comparing (\ref{LTB-R15}), (\ref{LTB-R2}) and (\ref{TB-3ugu}) we find out
that
\begin{eqnarray}
R(M,t) = \frac{9^{1/3}}{2}\,\left(\frac{4\,G\,M}{c^2}\right)^{1/3}
\left[t + \beta(M) \right]^{2/3},
\label{LTB-R1}
\end{eqnarray}
where
\begin{eqnarray}
\beta(M) = \frac{1}{3}\, \sqrt{\frac{2\,R_0^3(M)}{G\,M}}.
\label{new-beta}
\end{eqnarray}
Using the initial conditions (\ref{Mm}) we obtain
\begin{eqnarray}
\beta(M) = \sqrt{\frac{1}{6\,\pi\,G\,M}\,
\int\limits_{0}^{M}\frac{{\rm d}\,\tilde M}{\rho_0(\tilde M)}}.
\label{new-beta-1}
\end{eqnarray}

Bonnor (1974) studies one particular case of initial conditions
\begin{eqnarray}
\beta_0 = \beta(0) = 0.
\label{LTB-B}
\end{eqnarray}
It follows from (\ref{new-beta-1}) that
\begin{eqnarray}
\beta(0) = \frac{1}{\sqrt{6\,\pi\,G\,\rho_0(0)}}
\label{LTB-Bu}
\end{eqnarray}
so, (\ref{LTB-B}) means
\begin{eqnarray}
\rho_0(0) = \infty.
\label{LTB-Bubu}
\end{eqnarray}
At the same time it follows from (\ref{LTB-R1}) and (\ref{LTB-B}) that
\begin{eqnarray}
R(M,0) = 0,
\label{LTB-ubu}
\end{eqnarray}
so, there are no particles in the area $M > 0$.
This means that the initial condition (\ref{LTB-B}) produces the TB model
with delta-like distribution of dust:
\begin{eqnarray}
\rho_0(M) = \delta(M).
\label{i=ini}
\end{eqnarray}
The second example which we study is given by Ribeiro (1993).
Ribeiro puts
\begin{eqnarray}
F(r) = \alpha\,r^p, \qquad \beta(r) = \beta_0 + \eta_0\,r^q,
\label{LTB-Ra}
\end{eqnarray}
where $\alpha$, $\beta_0$, $\eta_0$, $p$ and $q$ are constants.
From (\ref{Bondi}) it follows that
\begin{eqnarray}
M = 4\,\pi\,\alpha\,r^p
\label{LTB-Ra1}
\end{eqnarray}
so,
\begin{eqnarray}
\beta(M) = \beta_0 + \eta_0\,\left(
\frac{M}{4\,\pi\alpha}
\right)^{q/p}.
\label{LTB-Ra2}
\end{eqnarray}
From (\ref{LTB-Ra2}) and (\ref{new-beta-1}) the equation for initial
density distribution it follows:
$$\rho_0(M) =$$
\begin{eqnarray}
\left\{
6\,\pi\,G\,
\left[
\beta_0 + \eta_0\,\left(
\frac{M}{4\,\pi\alpha}
\right)^{q/p}
\right]^2 +
\frac{12\,\pi\,G\,\,\eta_0\,q}{\left(4\,\,\pi\,\alpha\right)^{q/p}\,p}
\left[
\beta_0 + \eta_0\,\left(
\frac{M}{4\,\pi\,\alpha}
\right)^{q/p}
\right]
\,M^{q/p}
\right\}^{-1}.
\label{LTB-Ra3}
\end{eqnarray}
We obtain for $\rho_0(0)$:
\begin{eqnarray}
\rho_0(0) = \frac{1}{6\,\pi\,G\,\beta^2_0}.
\label{LTB-Ra44}
\end{eqnarray}
We also note that it follows from (\ref{LTB-Ra3}) that
\begin{eqnarray}
\left.\frac{\partial \rho_0(M)}{\partial M}\right|_{M = 0} = 0
\label{LTB-44}
\end{eqnarray}
in case of $q/p > 0$.

\section{Results}

The article studies the properties of the TB model with extra definition
of the transformation rule of the co-moving coordinates.

It is shown that the application of the definition (\ref{Bondi}) of
invariant mass as the transformation rule of co-moving coordinates allows
to separate the couples of functions $f$ and $F$ into the nonintersecting
classes and to fix the dependence $F$ on $f$ in every class.
It is shown that every class connected with the only system of coordinates
and characterized by the only function $\Psi$.
The Ruban-Chernin system of coordinates corresponds to the
identical transformation.

The Ruban-Chernin system of co-moving coordinates $\{M,t\}$ is used to
describe the TB model.

The functions $f$, $F$ and ${\bf F}$ are used in the TB model (Tolman 1934;
Bondi 1947) uses
as independent and undetermined. We have shown that the
transformation rule
(\ref{Bondi}) leads to the calssification of the set of couples of the
functions $(f,F)$ into nonintersecting classes, so the three undetermined
functions now are $f$, $\Psi$ and ${\bf F}$. The Ruban-Chernin system of
coordinates is marked by the equality $\Psi(M) = 1$, so only two functions
are undetermined: $f(M)$ and ${\bf F}(M)$.

Two forms of the initial conditions for the TB model are
presented in the section (\ref{s 5}): we can fix the
initial coordinate $R(M)$ and initial velocity
$\left.\frac{\partial R(M,t)}{\partial t}\right|_{t = 0}$
or, which is the same, the initial
profile of the density $\rho_0(M)$ and energy $E_0(M)$ distribution.
In the first case of initial conditions the function ${\bf F}(M)$ is
changed by the function $R_0(M)$ and the function $f(M)$ - by the
$\left.\frac{\partial R(M,t)}{\partial t}\right|_{t = 0}$.
In the second one the function ${\bf F}$
is changed by the function $\rho_0(M)$ and the function $f(M)$ by
the function $E_0(M)$.

The univalent definition of the Ruban-Chernin co-moving system of
coordinates allows to simplify the interpretation of the observations
using the model.

\section{Acknowledgements}

This article has been written under the influence of the results obtained
by Prof. William Bonnor, Prof. Arthur Chernin, Prof. Johne Moffat and Dr.
D.Tatarski.
I'm grateful for encouragement and discussion
to Prof. Arthur Chernin, Prof. Victor Brumberg, Prof. John Moffat,
Prof. Evgenij Edelman,
Dr.Yurij Baryshev, Dr. Andrzej Krasinski, Dr. Sergei Kopeikin, Dr.Sergei
Krasnikov, Dr.Roman Zapatrin and Marina Vasil'eva.
Dr. Baryshev has initiated my interest to the modern
Cosmology. Dr. Krasinski sent me his book "Physics in an Inhomogeneous
Universe". \newline
This paper was financially supported by "COSMION" Ltd., Moscow.

\medskip
\noindent{\bf\Large\bf References}
\medskip
{\small

Baryshev Yu.V., Sylos Labini F., Montuori M. \& Pietronero L. 1994,
Vistas in As\-
\newline $\mbox{  }$ $\mbox{  }$
tronomy, 38, 4.
\newline
%
Bondi H. 1947, MNRAS, 107, 410.
\newline
Bonnor W.B. 1972, MNRAS, 159, 261.
\newline
%
Bonnor W.B. 1974, MNRAS, 167, 55.
\newline
Bonnor W.B. \& Chamorro A. 1990, ApJ, 361, 1.
\newline
%
Chamorro A. 1991, ApJ, 383, 1.
\newline
%
Dwight H.B. 1961 "Tables of Integrals", NY.
\newline
Gromov A. 1996, gr-qc/9612038
\newline
%
Gurevich L.E. \& Chernin A.D. 1978, "The Introduction into Cosmogony",
Moscow,"Mir".
\newline
%
Kopeikin S. 1996, private communication.
\newline
%
Krasinski A. 1993, Physics in an inhomogeneous universe (a review),
Warszawa.
\newline
%
Landau L.D. \& Lifshits E.M. 1973, "The Field Theory", Moscow, "Nauka".
\newline
%
Moffat J.W. \& Tatarski D.S. 1992, Phys.Rev.D 45,10.
\newline
%
Moffat J.W. \& Tatarski D.S. 1995, ApJ, 453, 1.
\newline
%
Ribeiro M.B. 1992a, ApJ, 388, 1.
\newline
%
Ribeiro M.B. 1992b, ApJ 395, 29.
\newline
Ribeiro M.B. 1993, ApJ 415, 469.
\newline
%
Ruban V.A. \& Chernin A.D., 1969, The isotropisation of the nonhomogeneous
cos\-
\newline $\mbox{  }$ $\mbox{  }$
mological models. Proceeding of the 6th Winter School on the
Cosmophysics,
\newline $\mbox{  }$ $\mbox{  }$
p.15, Appatiti.
\newline
Tolman R.C. 1934, Proc.Nat.Acad.Sci (Wash), 20.
\newline
%
}

\end{document}